\documentclass[11pt, a4paper]{article}
\usepackage{amssymb, amsmath, amsbsy}
\usepackage{algorithmic, algorithm}
\usepackage{algorithm}
\usepackage{listings}
\usepackage[T1]{fontenc}
\usepackage{color}
\usepackage{authblk}
\usepackage{graphicx}
\usepackage{hyperref}

\title{Pushing a soft body droplet through porous medium}

\author[1]{Maciej Matyka \thanks{maciej.matyka@uwr.edu.pl}}
\affil[1]{Faculty of Physics and Astronomy, University of Wroc{\l}aw, pl.\ M.\ Borna 9, 50-205 Wroc{\l}aw, Poland, tel.: +48713759357, fax: +48713217682}
\begin{document}
\maketitle

\begin{abstract}
We investigate the problem of transport of a single fluid droplet through a non-wettable superhydrophobic porous medium.
A mechanical soft body model is developed and used to simulate the process of pushing fluid droplets through pore space of a random porous medium. 
Path lines of the center of mass of each droplet are used to calculate tortuosity of the media.
Our results show that droplet based tortuosity increases with decreasing porosity of the porous samples. Although qualitatively this agrees with the behaviour observed for tortuosity derived from the fluid flow, the form of this relation is different.
\end{abstract}


\section{Introduction}

The transport of individual fluid droplets in a porous medium is a subject of research on an in-pore organic matter mobilization and oil recovery \cite{Perazzo18} with a special focus on a single droplet trapping and mobilization \cite{Li05,He19}. Droplets motion at porous and textured surfaces rises many interesting scientific questions, e.g. the pancake bouncing phenomena \cite{Liu14, Liu16, Bro16} or the icing problem \cite{Remer14}. Imbibition of porous medium by a single droplet has important practical applications in the ink jet printing process \cite{Clarke02}.

Tortuosity $T$ is one of the key parameters in description of transport in porous media \cite{Clennell97}. It is a non-dimensional physical quantity that estimates the deviation of the transport paths from a straight line. One may define $T$ as the ratio of the average fluid flow path lengths $<\!\lambda\!>$ to system size $L$:
\begin{equation}
T=\frac{<\!\lambda\!>}{L}, \label{eq:tortuosity}
\end{equation}
where $<\!\lambda\!>$ is the effective length of the flow particles paths and $L$ is the porous medium length. 
We already know how tortuosity changes with porosity in the creeping flow regime of single-phase, incompressible fluid \cite{Koponen96,Matyka08}. $T$ may be defined for electric \cite{Zhang95} and diffusion or hydrodynamic transport processes as well \cite{Ghanbarian13, Saomoto15}. Due to the complexity of the multiphase flows, however, not much was done to investigate multiphase tortuosity in porous media.

Experimental methods for droplets are expensive, relatively inaccessible and time consuming. Thus, it is important to have tools for an efficient and flexible simulation of this phenomena. The motion of individual drops at pore-scale models of porous media were investigated before using i.e. the boundary-integral methods. This approach, however is inefficient and the simulation was restricted to maximum eighteen particles used to build porous media samples \cite{Davis09}. 
Also, the Lattice Boltzmann (LBM) models were used for simulation of the transport of separated fluid droplets \cite{Zhang19,Liu16}. However, controlling LBM multiphase simulations and, in particular, measuring quantities based on the single droplet movement is rather problematic as the usual output of the fluid flow solver is the fluid velocity and density fields. Thus, one has to extract droplets from these fields, identify them and track their motion in the porous medium. 

We aim at calculating multiphase, single droplet based tortuosity index. For this we will develop  the mechanical model of a fluid droplet based on the soft body model \cite{Matyka08, Matyka03}. We will simulate the transport of a single, non-wettable fluid through a complex, superhydrophobic porous medium build of randomly distributed grains. In our model, to calculate tortuosity we will track the path of the center of mass of each individual droplet. The model is efficient and allows us to push many droplets through the media. This allows extending the study to perform the statistical analysis of repeated numerical experiments. In particular, we will use it to calculate tortuosity versus porosity relation in a wide range of porosity and compare it to previous results obtained for single-phase flows \cite{Matyka08}.

\section{The Fluid Droplet Model}

\begin{figure}[!h]
\centering
\includegraphics[width=0.38\columnwidth]{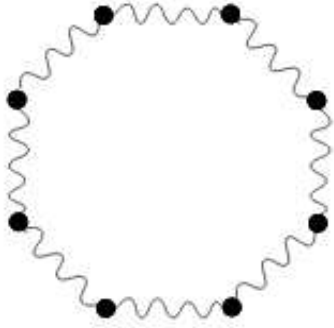}
\caption{The representation of a two-dimensional soft body model built from springs and masses. \label{springs}}
\end{figure}

To model fluid droplet transport in a porous medium we will assume that the medium is superhydrophobic (the wetting angle $180$ degree) and the droplet does not wet nor stick to the surface. Moreover, we assume the changes in its shape are small enough to ensure that the droplet will not separate into smaller ones. 
To model the droplet we will use the pressurized soft body model with the droplet built of springs and masses. Our previous work shows that this model is efficient in the simulation of soft objects that preserve shape and volume \cite{Matyka03}. 
To model the droplet we start with a set of masses distributed uniformly at its surface. Each mass is connected with its neighbor with a linear spring that represents the surface tension of the material (see Fig.~\ref{springs}. 
Three forces are acting on each mass point: gravity ($\vec{f}=m\cdot \vec{g}$), linear spring force between neighbors with damping:
\begin{equation}
\vec{f}_s= \left(-k_s (d - d_0) - k_d \frac{(\vec{v}_i-\vec{v}_j)\cdot\vec{r}_{ij}}{|\vec{r}_{ij}|}\right)\vec{r}_{ij}/|\vec{r}_{ij}|,
\end{equation}
where $k_s$ is the spring constant, $d$ is the distance between $i$-th and $j$-th mass, $d_0$ is the distance at rest, $k_d$ is the damping constant, $\vec{v}_i$ is the $i$-th mass velocity and $\vec{r}_{ij}$ is the difference between masses position $\vec{r}_i-\vec{r}_j$. If not stated otherwise, we used $\vec{g}=(0,-5\cdot10^5)$, $k_s=119755$ and $k_d=365$ in numerical units.

To model the fluid inside the droplet, we neglect the fluid dynamics inside and assume constant pressure across the interface. Thus, the third force  appears in the model and acts on the surface due to the difference between internal and atmospheric pressure. To compute its value we use the Ideal Gas Law:
\begin{equation}
    PV=nRT,
\end{equation}
where $P$ is the pressure, $R$ is the gas constant, $n$ is the number of moles and $T$ is the gas temperature and $V$ is the volume of the droplet. We assume $P$, $R$, $n$ and $T$ are constants. The instant volume during the simulation at each computational step is only needed and for this, we use the Gauss theorem and decrease the dimension of the problem by one. For example, in two dimensions, the surface field may be obtained as an integral over the boundary and approximated as:
\begin{equation}
S = \int \int_S dS=\oint_l x n_x dl\sum^{N_L}_{i=1}x_in_{x,i}l_i, \label{eq:volume}
\end{equation}
where $S$ is the surface of the droplet, $dS$ is an infinitesimal element on the surface, $l$ is the droplet boundary, $N_L$ is the number of boundary segments, $x$ denotes the position in the $x$ direction, $n_x$ is the $x$ component of the normal vector, $dl$ is the infinitesimal element of the droplet boundary, $n_{x,i}$ is the normal vector to the $i-th$ boundary segment, $l_i$ is the length of the $i-th$ segment.

Having computed $V$ from equation (\ref{eq:volume}), we may now compute the pressure force using the normal vector $\hat{n}$ to the surface:
\begin{equation}
\vec{f}_p = \frac{nRT}{V} \hat{n}.
\end{equation}

After calculating the forces, we sum them up for each mass point and integrate equations of motion. For integration, we used the second-order Verlet algorithm and with $\delta t=2\cdot 10^{-5}$ with additional inverse dynamic constraints that stabilized the simulation \cite{Provot95}.

\section{Collisions}

We model the porous medium as a system of 2d, random, overlapping and non-overlapping impermeable grains (discs) of varying radius. By varying the number of grains we control the porosity. In the non-overlapping case we kept the minimal distance between grains at 13$\%$ of their radius to prevent them from touching. For collisions of the soft body with solid grains, we used the penalty method. Each time any of the masses at the surface of the droplet wants to penetrate any of the solid grains, we add a virtual spring that has $0$ resting length and acts outwards the grain surface. A similar technique was previously used i.e. to model the static friction in granular material simulations\cite{Risto94}.
Our collision approach is, thus, similar to Hertz type of collision models, as we also assume some degree of softness of the surface.
\begin{figure}[!h]
\centering
\includegraphics[width=0.71\columnwidth]{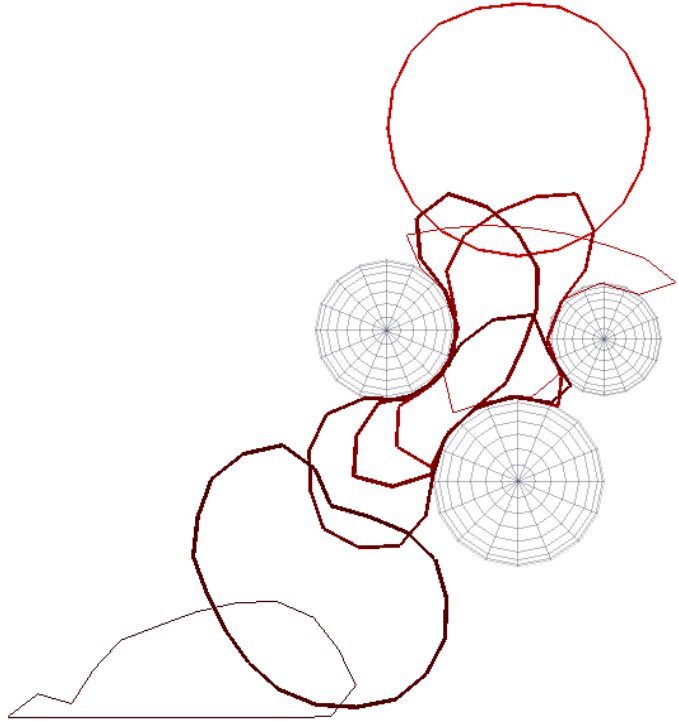}
\caption{Exemplary simulation of a fluid drop that pushes through the single pore created by three solid grains. We plot selected time snapshots of the evolving droplet's boundary \label{fig:collision}.}
\end{figure}
As shown in Fig.~\ref{fig:collision} the fluid droplet which starts just above the solid grains (round shape at the top) and goes down under gravity, changing its shape on its way as the passage becomes narrow.

\section{Results}

To study the tortuosity-porosity relation we generate random configurations consisting of overlapping grains. By varying the number of grains we varied porosity of the system. At each porosity, $200$ independent configurations of grains were used. For each configuration, maximum of $200$ droplets were simulated. 
The actual number was dependent on the number of successful simulations, because in low porous systems we faced the problem of blocking of droplets in dead-end pores \cite{Andrade97} and, thus, the simulation was time-consuming and sometimes did not converge.
For each simulation, the droplet was placed randomly just upper edge of the porous medium starts.

\begin{figure}[!h]
\centering
\includegraphics[width=0.75\columnwidth]{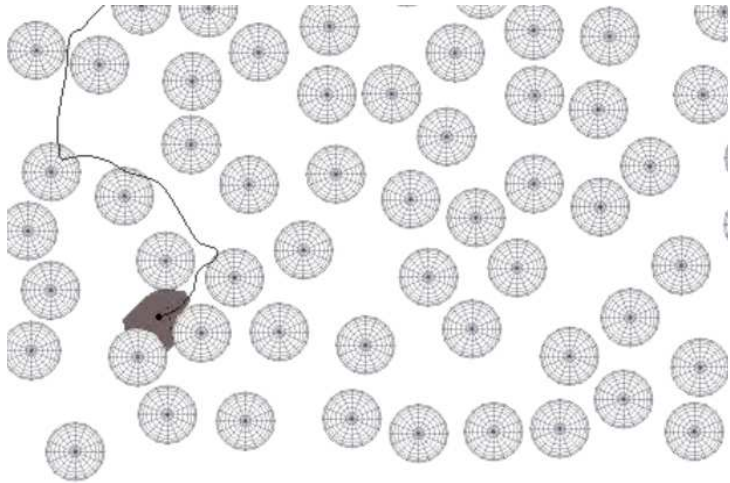}
\caption{Animation still frame showing real-time simulation of pushing of the soft body droplet through overlapping and non-overlapping porous medium in gravity field. The animation can be viewed here: \url{http://youtu.be/KwXKz9zCR9Y}\label{fig:animation}}
\end{figure}

\begin{figure}[!h]
\centering
\includegraphics[width=0.75\columnwidth]{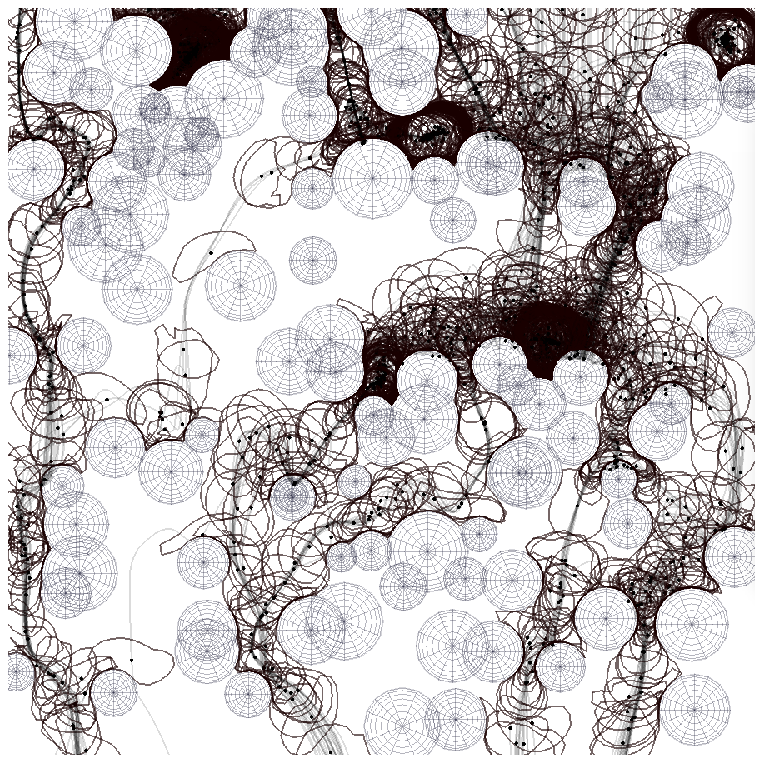}
\caption{The snapshot of the time history of a few individual droplets (the center of mass drawn as a solid black dot) pushed through the porous model built of random overlapping, solid disks (wire spheres in the visualization).\label{fig:timesnapshots}}
\end{figure}

The exemplary paths and time snapshots of simulations of the droplets pushed through one of the porous samples is given in Fig.~\ref{fig:timesnapshots}. 
The corresponding animation file is available in suplementary material and can be viewed in the electronic version of the article, see Fig.~\ref{fig:animation}.

\begin{figure}[!h]
\centering
\includegraphics[width=0.85\columnwidth]{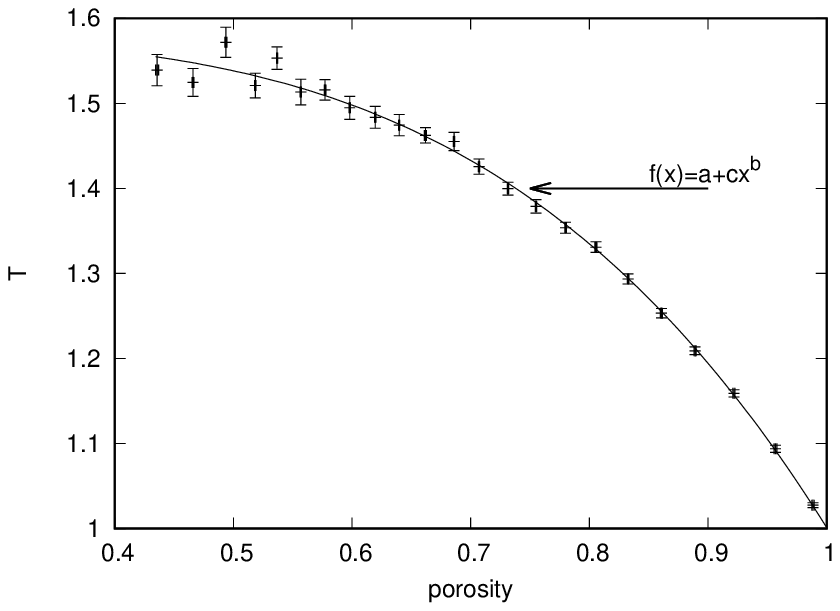}
\caption{The average tortuosity in the droplet model at varying porosity. Each point represents the average tortuosity and each error bar is the standard error based on the $200$ independent configurations of obstacles. The solid line is the best fit to the function $f(x)=a-cx^b$. We found $a=1.58$, $b=3.87$ and $c=0.58$ using the least square fitting algorithm. \label{fig:tortuosity}}
\end{figure}

The visualization technique uses the alpha blending, thanks to which one may observe that some droplets get stuck, e.g. in the dark region in the left top gets colored because many droplets stay in this region forever). We may also identify that some parts of the porous matrix are relatively more permeable and let many droplets go through (top-right part of Fig.~\ref{fig:timesnapshots}).
Using the stored paths for all droplets in all porous samples that we simulated, we used the path-based definition of tortuosity, Eq.~\ref{eq:tortuosity}, and plotted the resulting average tortuosity versus porosity in Fig.~\ref{fig:tortuosity}.
By numerical fitting of various $T(\varphi)$ relations we found that the best match is obtained with the power law:

\begin{equation}\label{fittingt}
    f(x)=a-cx^b,
\end{equation}
(see Fig.~\ref{fig:tortuosity} for fitting data).

\begin{figure}[!h]
\centering
\includegraphics[width=1.0\columnwidth]{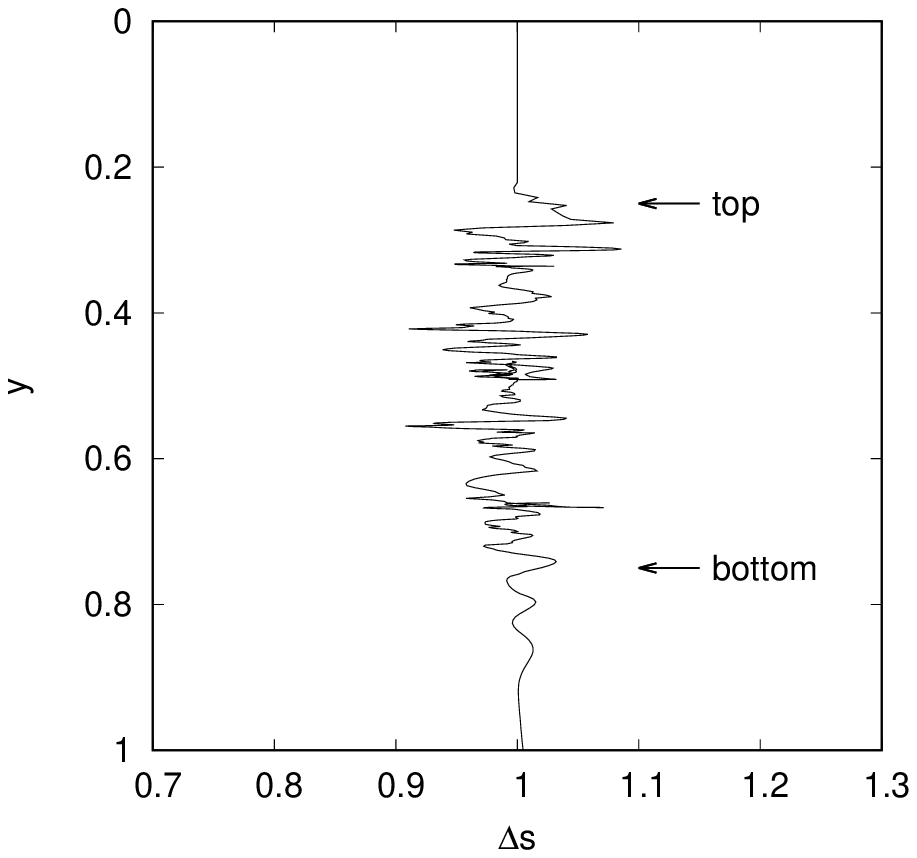}
\caption{The surface change $\Delta S$ for a single droplet simulation in the $\varphi=0.79$ porous medium build of non-overlapping and slightly separated grains. Arrows show the position of the top and bottom interface of the medium (the droplet, initially placed above the medium upper edge was pushed through the medium and then continued the motion after getting out of it through the floow).\label{fig:resultssingle}}.
\end{figure}

Next, we adopted our model to investigate the motion of a single droplet in the porous matrix build of non-overlapping grains (this change made the model more permeable at lower porosity). In particular, we were interested in the volume changes and, thus, we monitored the change of droplet superficial area $\Delta S=S_p(t)/S_0$, where $S_p(t)$ is the droplet area in a porous medium that change over time as the droplet pushes through the porous matrix and $S_0$ is the area before the porous matrix was hit. Our results are given in Fig.~\ref{fig:resultssingle}.

We observe the qualitative agreement of our results with simulations and experiments reported in \cite{Zhang19}. In our case, the droplet both shrinks and expands as it goes through the media. However, in \cite{Zhang19} most of the simulations that show the lower limit of $\Delta S=1$ and $\Delta S < 1$ was obtained for a special case of two droplets (one pushed after another). The modeling of two droplets is out of the scope of this paper but deserves more attention in the future.

\begin{figure}[!h]
\centering
\includegraphics[width=0.85\columnwidth]{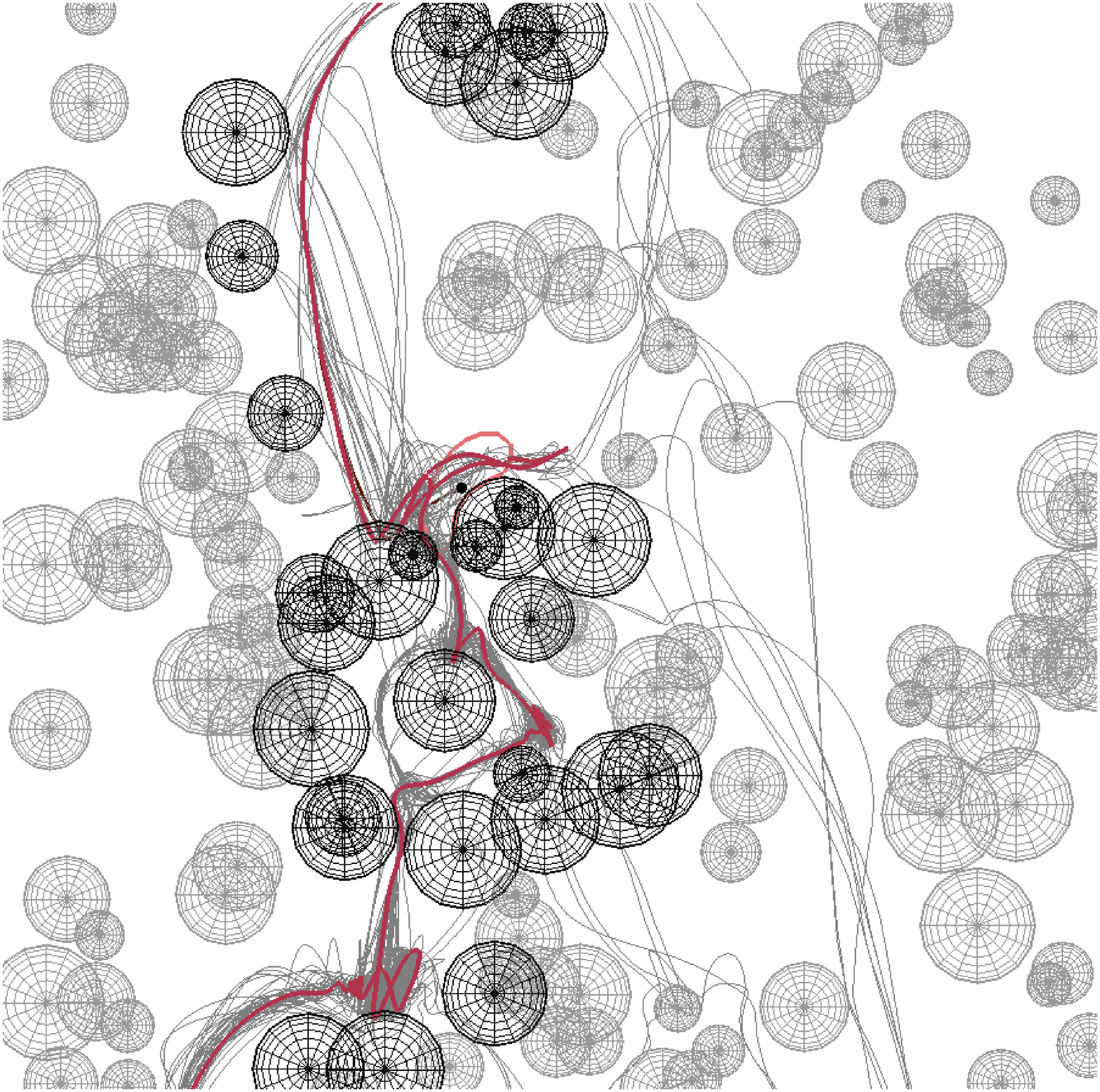}
\caption{The genetic algorithm seeks for the configuration of maximum tortuosity. Here, we found the system of $T=2.42$. The solid thick line is the visualization of the longest path. The thin lines represent the set of generated paths of lower tortuosity. The instant configuration at the highest tortuosity is represented by disks visualized as wire spheres. Dark spheres are those "touched" by the droplet in the previous simulation. \label{fig:genetic}}
\end{figure}

Finally, we have used our model to artificially generate the system at maximum tortuosity. Here, we utilized the fact that the simulation is very fast and this allowed us to calculate thousands of repetitions.
The algorithm was as follows. First, we generate a random configuration of some fixed number of grains. Then we simulate the pushing process starting at half of the sample just above it (see Fig.~\ref{fig:genetic}).
From this we determine the tortuosity at the current configuration. 
After the simulation we slightly change the position of the grains and run the simulation again. We accept and use the new configuration only if the tortuosity has increased.

To probe the wide range of possible configuration space we do the following: normally the position change procedure slightly and randomly changes the position of all grains in our sample. However, each 10-th time the special procedure is called. During each simulation we keep the list of grains that are touched by the droplet. In this special case we use this information and slightly change the position of one of these touched grains. This directly influences the path of the droplet in the next simulation and is expected to change the tortuosity as well. 

\section{Discussion and Conclusions}

Our results confirm previous findings based on transport of single-phase fluid flows where tortuosity grows monotonically with decreasing porosity in the same type, random overlapping obstacles, porous medium model \cite{Koponen96,Matyka08}. This suggest that in the single phase fluids as well as in the multiphase, individual fluid droplets the largest, most representative pores are chosen as the main channels for the transport. 
In \cite{Matyka08}, the logarithmic tortuosity law 
\begin{equation}
T(\varphi)=1-p\log(\varphi)
\end{equation}
was found as the best approximation in the single-phase fluid flow through overlapping quads porous medium model. 
Here, using the fitting procedure we found, that the power law holds:
\begin{equation}
T(\varphi)-T_0 \propto c\varphi^b, \label{powerlaw}
\end{equation}
where $T_0=1.58$ is the maximum possible value of averaged tortuosity taken close to the percolation threshold. $c$ and $b$ are empirical constants depend the model. Similar power law tortuosity versus porosity relations were previously found, e.g. in \cite{Araujo06} where $T\propto \varphi^{-2}$ was suggested for fluid flow through diluted porous medium build of non touching (distanced) solid grains. None of the above mentioned works was related to single droplet movement. In this context our result is original and it may be worth further investigation to explain why the power law holds in tortuosity-porosity relations among various transport phenomena.

Our result may be related to the Archie Law describing electrical and fluid transport through porous medium:
\begin{equation} 
\label{Archie}
T=\varphi^{-b},
\end{equation}
where i.e. $b\approx 0.25$ for three dimensional overlapping spheres model \cite{Matyka12}. In the droplet transport, however, the initial shifting of $T$ was necessary and the power law exponent, thus, differ. 
We have also found that, because of the way the data are averaged, one may obtain various types of scaling. 
\begin{figure}
    \centering
    \includegraphics[width=0.95\columnwidth]{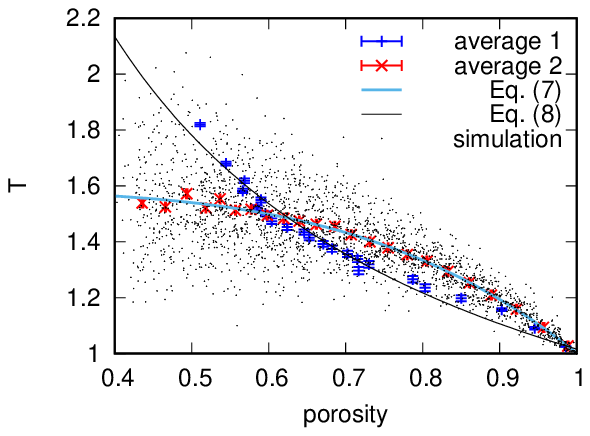}
    \caption{(color figure) Droplet tortuosity measured in random porous medium. Each small black dot represent one porous medium and $200$ independent droplets pushed through the medium. Point data with the error bars represent two types of averaging taken over these data.}
    \label{fig:tortuosityanalysis}
\end{figure}
In Fig.~\ref{fig:tortuosityanalysis} we have plotted the original tortuosity measurements (small, black dots) with two types of average plots. Each point in these averages is computed from $100$ points selected from the original measurements. However, in the average type 1 the results were initially sorted by tortuosity, whereas in average 2 they were sorted by porosity. It is stunning how different these plots are, especially that they are taken from the same data set. It shows how important it is to do a proper statistics of the data that are randomly generated. For example, it may be possible that we will obtain a simple power law for the hydrodynamic tortuosity just, if we initially sort the original data in respect to porosity \cite{Matyka08}. The further analysis oif hydrodynamic tortuosity is, however, out of the scope of this paper.


Technically, from the droplet model point of view, one of the key issues we had during the simulation was the blocking droplets in the low porous systems. Thus, for the simulation of pushing the droplet in low porous systems and surface changes investigation we had to change the model from overlapping to the non-overlapping. This significantly improved the simulation. One may also consider to include some kind of vibration mechanism of dealing with trapped droplets (i.e. vibrations \cite{Li05}) which may be implemented in the presented model easily. 

\section{Acknowledgments}

I would like to thank prof. Zbigniew Koza from Faculty of Physics and Astronomy at University of Wroc{\l}aw for reading the first version of the manuscript and for his valuable comments how to improve it.
This work was supported by grant 
NCN Miniatura 2 Nr 2018/02/X/ST3/03075.

\bibliographystyle{unsrt}  

\end{document}